# Crystallinity Evolution of MOCVD-Grown β-Ga$_2$O$_3$ Films Probed by *in situ* HT-XRD under Different Reactor Heights

Imteaz Rahaman[1,*], Botong Li[1,*], Bobby G. Duersch[2], Hunter D. Ellis[1], Kathy Anderson[3], and Kai Fu[1, a)]

[1]Department of Electrical and Computer Engineering, The University of Utah, Salt Lake City, UT 84112, USA

[2]Utah Nanofab Electron Microscopy and Surface Analysis Laboratory, The University of Utah, Salt Lake City, UT 84112, USA

[3]Utah Nanofab, Price College of Engineering, The University of Utah, Salt Lake City, Utah 84112, United States

**Abstract**

The crystallinity of β-Ga$_2$O$_3$ thin films grown by metal–organic chemical vapor deposition (MOCVD) is strongly influenced by reactor design and the resulting growth environment. In this work, we investigate the role of reactor height on the crystallinity evolution of MOCVD-grown β-Ga$_2$O$_3$ films by directly comparing long- and short-chamber showerhead configurations. Structural evolution was probed by *in situ* high-temperature X-ray diffraction (HT-XRD) as the MOCVD-grown films were heated from 25 °C to 1100 °C. Temperature-dependent XRD reveals a consistent redshift of the β-Ga$_2$O$_3$ (–201) reflection after HT-XRD heating and subsequent cooling to room temperature for both reactor geometries, indicating a similar thermally driven strain response. Quantitative rocking-curve analysis shows a non-monotonic temperature dependence of the (–201) full width at half maximum (FWHM), with minimum values of approximately 2.03° and 2.72° for the short- and long-chamber films, respectively, reflecting differences in mosaic alignment established during growth. Atomic force microscopy further shows that short-chamber-grown films exhibit smoother surfaces, with root-mean-square roughness values of approximately 7.7 nm before and 7.3 nm after HT-XRD heating, compared to 19.3 nm and 12.3 nm, respectively, for long-chamber-grown films. Overall, these results indicate that reactor heights influence the initial crystalline and morphological templates of β-Ga$_2$O$_3$ films and modulate their elevated-temperature structural response, providing practical insights for optimizing MOCVD reactor design for high-quality β-Ga$_2$O$_3$ growth.

———————————————
* These authors contributed equally to this work.

a) Author to whom correspondence should be addressed. Electronic mail: kai.fu@utah.edu

Ultra-wide bandgap (UWBG) semiconductors have emerged as key enablers for next-generation power electronics and harsh-environment applications due to their ability to support high breakdown fields, operate at elevated temperatures, and sustain large power densities [1–3]. Among them, β-Ga$_2$O$_3$ has attracted considerable attention owing to its large bandgap (~4.8–4.9 eV), exceptionally high theoretical critical electric field (>8 MV cm$^{-1}$), and the availability of large-area melt-grown substrates[4–6]. These attributes make β-Ga$_2$O$_3$ a promising platform for power rectifiers, field-effect transistors, solar-blind photodetectors, and radiation-tolerant electronics [7–10]. However, the practical performance of β-Ga$_2$O$_3$ devices remains limited by crystalline defects, mosaic spread, and surface roughness introduced during thin-film growth, particularly for heteroepitaxial films grown on lattice- and symmetry-mismatched substrates [11–13]. MOCVD is widely employed for β-Ga$_2$O$_3$ thin-film growth due to its scalability, precise control of growth parameters, and compatibility with industrial manufacturing [14,15]. Previous studies have demonstrated that growth temperature, chamber pressure, precursor chemistry, oxygen partial pressure, and substrate selection critically influence film crystallinity and morphology[16–21]. Homoepitaxial β-Ga$_2$O$_3$ films typically exhibit narrow X-ray diffraction rocking-curve linewidths and low defect densities, whereas heteroepitaxial growth on *c*-plane sapphire, GaN, MgO, or GaAs often results in increased mosaicity and surface roughness due to lattice mismatch and symmetry incompatibility[22–25]. Post-growth thermal annealing has been extensively used to improve crystalline ordering through defect annihilation and strain relaxation [26–29]. However, reported annealing outcomes vary widely across the literature, even under nominally similar thermal conditions, indicating that the annealing response is strongly dependent on the as-grown film quality and growth environment. Notably, while extensive efforts have focused on optimizing growth parameters, the influence of MOCVD reactor geometry—particularly chamber height and its impact on gas-phase transport and residence time—has received limited systematic investigation.

In this work, we demonstrate that reactor ceiling height is a critical and previously underappreciated factor governing the crystalline evolution of MOCVD-grown β-$Ga_2O_3$ films during *in situ* HT-XRD heating. By systematically comparing long- and short-chamber showerhead MOCVD configurations, we isolate the role of chamber height in controlling precursor transport, gas residence time, and the subsequent structural response to thermal treatment. Using comprehensive X-ray diffraction analysis, rocking-curve measurements, phi-scan characterization, atomic force microscopy, and benchmarking against reported literature, we establish clear correlations between reactor height, annealing behavior, crystalline quality, and surface morphology. This study provides new insights into how reactor-level design parameters influence β-$Ga_2O_3$ film quality beyond conventional growth-condition optimization and offers practical guidance for tailoring MOCVD reactors and post-growth processing strategies for high-quality β-$Ga_2O_3$ epilayers relevant to UWBG electronic and optoelectronic device applications.

The β-Ga$_2$O$_3$ epitaxial layers were grown on α-Al$_2$O$_3$ (0001) substrates without intentional off-cut using MOCVD. Prior to growth, the substrates were immersed in piranha solution for 10–15 min to remove organic contaminants, and all pre-growth preparation steps were conducted at room temperature. The depositions were carried out in a customized vertical-type showerhead MOCVD reactor manufactured by Agnitron Technology, similar to that used in our previous studies.[30–32] To ensure thickness uniformity, the susceptor rotation speed and chamber pressure were maintained at 150 rpm and 60 Torr, respectively, for all experiments. The temperature ramping procedure consisted of two stages: the chamber was first heated linearly to 750 °C and held for 4 min, followed by a second linear ramp to the final growth temperature of 850 °C. During epitaxy, triethylgallium (TEGa) and high-purity O$_2$ were used as the Ga and O precursors, respectively, with N$_2$ as the carrier gas. The flow rates of TEGa and O$_2$ were fixed at 130 sccm and 800 sccm, respectively. The growth duration was 40 min for all samples. HT-XRD measurements were performed using a Bruker D8 DISCOVER high-resolution X-ray diffractometer equipped with a Cu K$\alpha_1$ source ($\lambda$ = 1.5406 Å), a triple-bounce channel-cut monochromator, and an Anton Paar graphite dome hot stage. The HT-XRD scans were conducted in air at temperatures from 25 °C up to 1100 °C, followed by cooling to room temperature. Atomic force microscopy (AFM) measurements were carried out in tapping mode using a Bruker Dimension ICON AFM, with all images acquired over a 5 × 5 μm$^2$ scan area.

Figure 1 illustrates the schematic of the showerhead-type MOCVD reactor employed in this study and highlights the geometric distinction between the long- and short-chamber configurations through the ceiling height $H_C$, defined as the vertical distance between the showerhead and the substrate. In both configurations, metal–organic precursors and oxidant gases are delivered through independent gas lines and introduced via a perforated showerhead, establishing a near-stagnation flow field above the substrate mounted on an RF-heated susceptor. Showerhead reactors are widely used in MOCVD because they enable laterally uniform gas distribution and decouple gas-phase transport from surface reactions. From a mass-transport perspective, the ceiling height, $H_C$ plays a central role in determining the gas residence time and the relative contributions of diffusion and convection to precursor transport. The gas residence time is given by [33,34]

$$t_{res} = \frac{V}{F_{in}} \quad (1)$$

where '$V$' is the effective reactor volume and '$F_{in}$' is the inlet volumetric flow rate. Since the chamber volume scales approximately with $H_C$, variations in ceiling height directly modify $t_{res}$. In addition to residence time, the effective consumption time associated with surface reactions can be expressed as [33,34]

$$t_{con} = \frac{V}{K_S S} \quad (2)$$

where '$K_S$' is the surface reaction rate constant and '$S$' is the growth surface area. The relative magnitude of $t_{res}$ and $t_{con}$ determines whether growth proceeds in a surface-reaction-limited regime ($t_{res} \ll t_{con}$) or a mass-transport-limited regime ($t_{con} < t_{res}$). The characteristic diffusion length for reactive species can be expressed as [33,34]

$$L_d = \sqrt{4Dt} \quad (3)$$

where $D$ is the gas-phase diffusivity and $t$ is the characteristic transport time, often comparable to $t_{res}$. The ratio between $L_d$ and $H_C$ governs whether vertical transport is dominated by diffusion or influenced by convective effects, while also affecting the degree of lateral redistribution prior to surface incorporation. For showerhead reactors, the effective residence time experienced by species approaching the substrate may further include diffusion across the boundary layer and can be written as [33,34]

$$t' = \frac{H_C}{v_{in}} \ln\left(\frac{\delta}{H_C}\right) + \frac{\delta^2}{4D} \qquad (4)$$

where '$v_{in}$' is the inlet gas velocity and '$\delta$' is the boundary-layer thickness. This expression highlights that the effective residence time scales approximately linearly with $H_C$, reinforcing the role of reactor height as a critical control parameter. Reducing the ceiling height decreases the residence time and diffusion length, thereby suppressing gas-phase reactions and limiting precursor depletion in systems where parasitic chemistry is dominant. However, shorter residence time may also reduce gas-phase homogenization and increase sensitivity to local fluctuations in precursor concentration and temperature, particularly in growth regimes where surface reaction kinetics are rapid and mass transport limits incorporation. Conversely, increasing the ceiling height increases $t_{res}$, allowing greater gas-phase equilibration and stabilization of concentration boundary layers, but may also enhance unwanted gas-phase reactions depending on precursor chemistry and growth conditions.

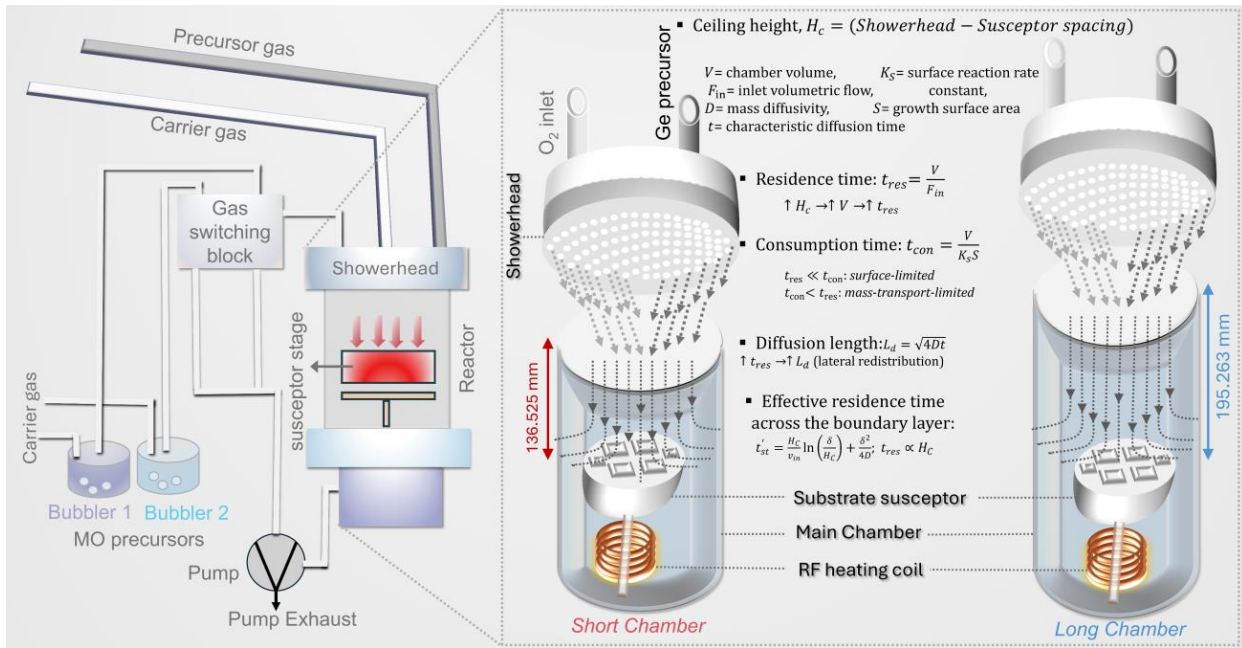

**FIG. 1.** Schematic of the showerhead MOCVD reactor and comparison between long- and short-chamber configurations. The left panel shows the overall reactor layout, including precursor

delivery, showerhead, susceptor, and exhaust. The right panels illustrate the short-chamber (136.525 mm) and long-chamber (195.263 mm) geometries, highlighting the difference in showerhead–susceptor spacing and its impact on gas residence time and precursor transport to the substrate.

As a result, shortening the ceiling height does not necessarily improve film quality. Instead, the optimal chamber height depends on the balance between gas-phase reactions, diffusion-dominated transport, and surface kinetics specific to a given material system. In the following sections, we experimentally evaluate how these competing effects manifest in β-$Ga_2O_3$ growth by correlating reactor height with crystalline quality and annealing behavior using X-ray diffraction and surface-morphology analyses.

Figure 2 shows the structural evolution of β-$Ga_2O_3$ films grown using long- and short-chamber MOCVD configurations as a function of post-growth *in situ* HT-XRD heating. Figures 2(a) and 2(c) present the full *2θ–ω* XRD diffractograms, while Figs. 2(b) and 2(d) display enlarged views of the β-$Ga_2O_3$ (–201) reflection to track changes in peak position with HT-XRD heating. For both long- and short-chamber-grown films, the β-$Ga_2O_3$ diffraction peaks are present in the as-grown state and persist throughout the HT-XRD heating sequence, indicating that the crystalline phase remains stable during thermal treatment. Within the measurement resolution, no pronounced or systematic changes in peak intensity or apparent peak sharpening are observed as a function of HT-XRD heating temperature in either dataset. A consistent shift of the β-$Ga_2O_3$ (–201) reflection toward lower *2θ* values is observed after HT-XRD heating for both films. This red shift indicates an increase in the out-of-plane lattice spacing and is attributed to partial relaxation of residual growth-induced strain. HT-XRD may facilitate defect redistribution and stress-relaxation mechanisms, such as point-defect reconfiguration and dislocation rearrangement, leading to similar lattice-parameter evolution in both films. The comparable magnitude and direction of the peak shift suggest that the strain-relaxation process is intrinsic to β-Ga$_2$O$_3$ and is largely independent of reactor height.

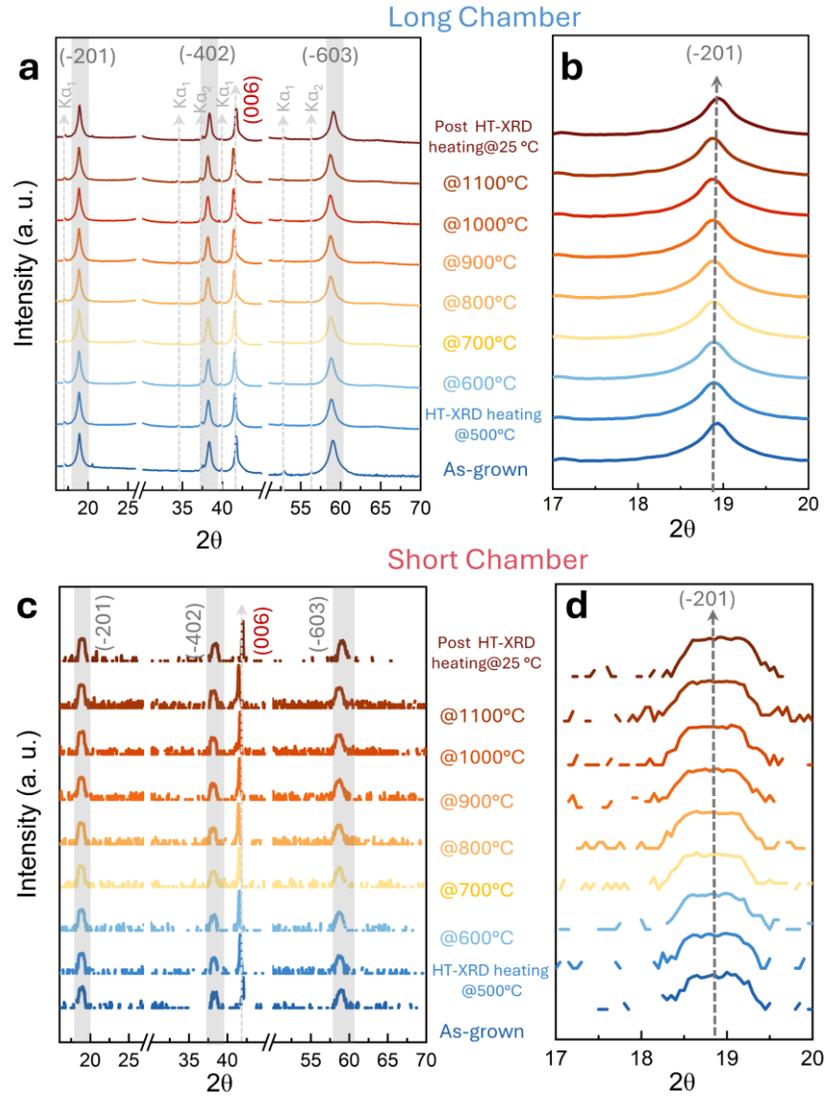

**FIG. 2.** Temperature-dependent X-ray diffraction of β-Ga$_2$O$_3$ films grown in long- and short-chamber MOCVD configurations. Diffraction patterns were collected using *in situ* HT-XRD on MOCVD-grown films from room temperature to 1100 °C in air. (a,b) 2θ–ω scans and magnified views of the β-Ga$_2$O$_3$ (–201) reflection for the long-chamber film. (c,d) Corresponding data for the short-chamber film. The dashed line marks the reference position of the β-Ga$_2$O$_3$ (–201) reflection, highlighting the redshift toward lower *2θ* values observed after HT-XRD heating for both films.

Figure 3 provides a quantitative assessment of the out-of-plane and in-plane crystalline characteristics of β-Ga$_2$O$_3$ films grown in long- and short-chamber MOCVD configurations using ω-scan and φ-scan XRD measurements.

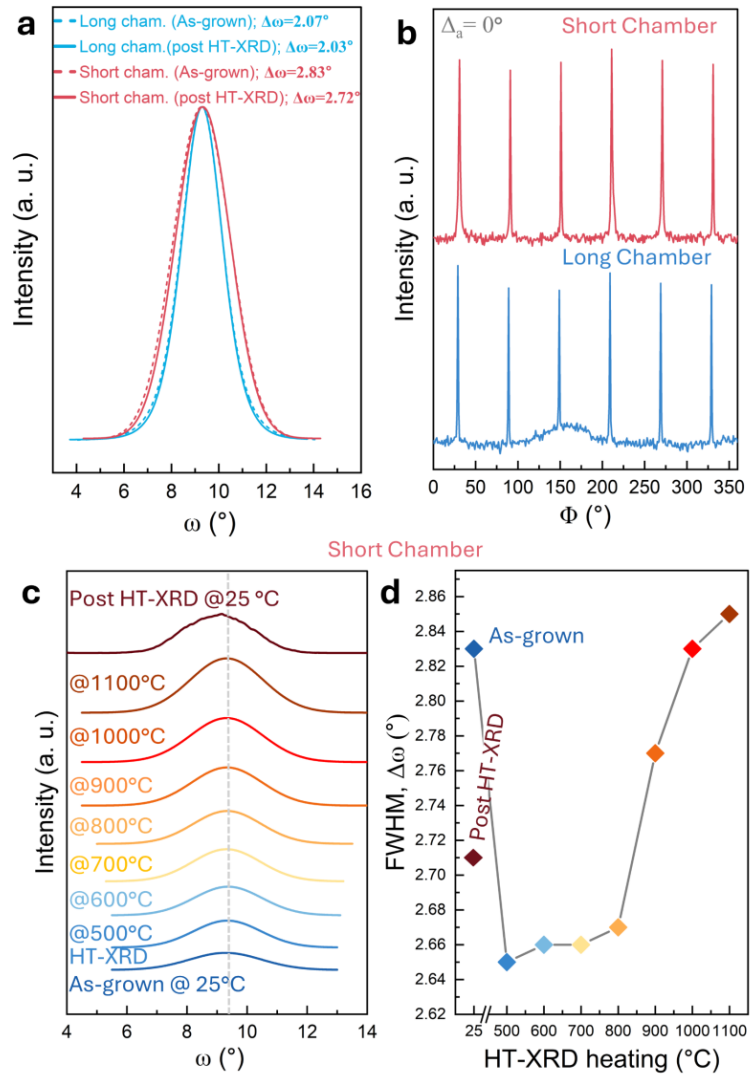

**FIG. 3.** Rocking-curve and in-plane orientation analysis of β-Ga$_2$O$_3$ films grown in long- and short-chamber MOCVD configurations. Data were collected during HT-XRD measurements on MOCVD-grown films, heating from 25 °C to 1100 °C. (a) ω-scan profiles of the β-Ga$_2$O$_3$ (–201)

reflection measured for long- and short-chamber films at room temperature and after HT-XRD heating. (b) φ-scan of the β-Ga$_2$O$_3$ (113) reflection collected after HT-XRD heating for both films, showing periodic in-plane diffraction features. (c) Evolution of the β-Ga$_2$O$_3$ (–201) ω-scan profiles for the short-chamber film as a function of measurement temperature during HT-XRD. (d) FWHM of the (–201) reflection as a function of measurement temperature, illustrating a non-monotonic temperature dependence of mosaic alignment.

Figure 3(a) compares the ω-scan (rocking-curve) profiles of the β-Ga$_2$O$_3$ (–201) reflection for films grown in both reactor geometries, measured at room temperature and after *in situ* HT-XRD heating. In the as-grown state, the short-chamber film exhibits a larger FWHM than the long-chamber film, indicating a higher degree of mosaic spread. When considered within the mass-transport framework introduced in Figure 1, this difference may originate from variations in the initial growth environment. Changes in gas residence time and diffusion length can alter precursor transport near the substrate, potentially influencing the mosaic structure formed during deposition. Upon *in situ* high-temperature XRD heating, changes in the rocking-curve profiles are observed for both films, indicating that crystalline alignment is sensitive to elevated-temperature exposure. The subsequent temperature-dependent response appears to act on the as-grown structural template, yielding broadly comparable trends without evidence of height-selective behavior. Overall, these observations suggest that HT-XRD provides a sensitive, quantitative probe of mosaic evolution in β-Ga$_2$O$_3$ films, while highlighting the coupled—and possibly interdependent—influence of growth conditions and elevated-temperature structural response.

Figure 3(b) shows φ-scan measurements of the β-Ga$_2$O$_3$ (113) reflection collected after high-temperature XRD heating for both films. Periodic diffraction peaks separated by 60° are observed in both cases, indicating the presence of multiple in-plane rotational domains. This behavior is consistent with prior reports[35] on heteroepitaxial β-Ga$_2$O$_3$ films grown on α-Al$_2$O$_3$, where the monoclinic symmetry of β-Ga$_2$O$_3$ combined with the higher symmetry of the substrate can give rise to rotational domain formation. The similar angular periodicity observed for both reactor configurations suggests that elevated-temperature XRD heating did not fundamentally alter the in-plane domain structure. Figures 3(c) and 3(d) show the evolution of the ω-scan profiles and extracted FWHM values for the short-chamber-grown film as a function of measurement temperature during HT-XRD. FWHM exhibits a non-monotonic dependence on temperature, initially decreasing and then increasing at higher temperatures. This behavior may result from competing thermally activated processes, where partial strain relaxation or defect reconfiguration possibly reduces mosaic spread at intermediate temperatures, while higher temperatures may activate additional lattice distortion or defect generation.

Figure 4 compares the surface morphology of β-Ga$_2$O$_3$ films grown in long- and short-chamber MOCVD configurations using atomic force microscopy (AFM). Figures 4(a) and 4(b) show the surface topography of the films before HT-XRD heating, while Figs. 4(c) and 4(d) present the corresponding morphologies after HT-XRD heating. Before HT-XRD heating, the long-chamber-grown film [Fig. 4(a)] exhibits a rough and non-uniform surface, with a relatively high root-mean-square (RMS) roughness of ~19.25 nm. The surface is characterized by pronounced height variations and localized protrusions. In contrast, the short-chamber-grown film [Fig. 4(b)] displays a comparatively smoother and more continuous morphology, with a lower RMS roughness of ~7.74 nm. This difference may reflect variations in surface diffusion and nucleation behavior established during growth under different reactor geometries. After HT-XRD heating, the long-chamber film [Fig. 4(c)] shows a reduction in RMS roughness to ~12.31 nm, suggesting partial surface relaxation or redistribution of surface features during elevated-temperature exposure. However, noticeable height variations and surface inhomogeneity remain. The short-chamber film [Fig. 4(d)] retains a relatively smooth morphology after heating, with only a modest change in RMS roughness (~7.29 nm), indicating that elevated-temperature exposure does not strongly perturb the surface topography in this case.

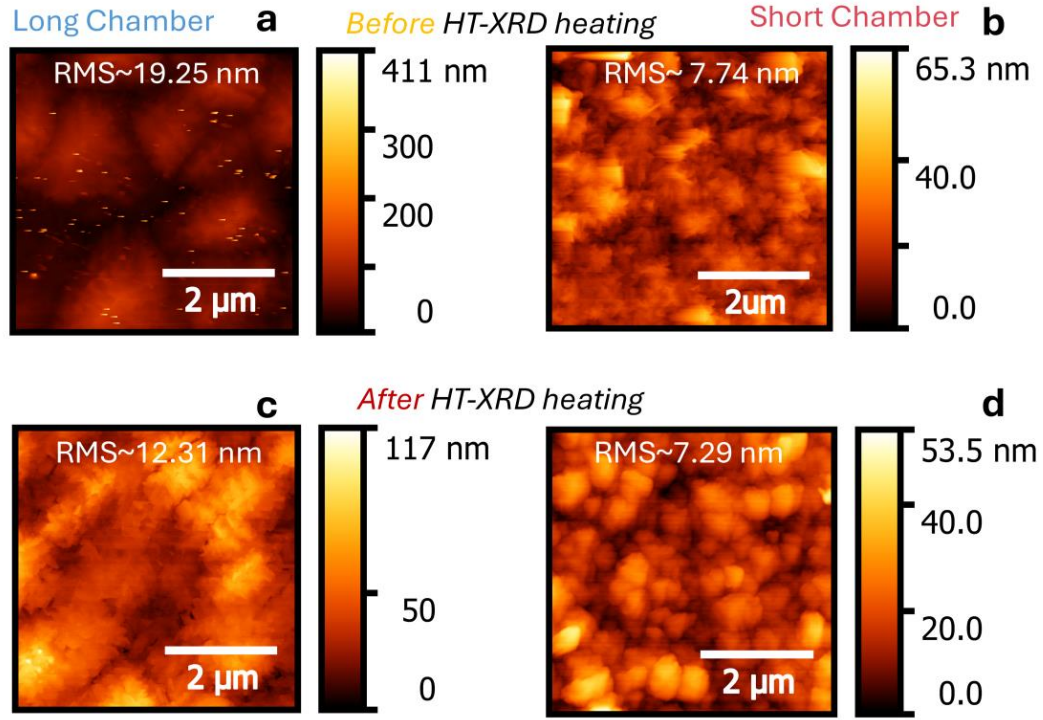

**FIG. 4.** AFM surface morphology of β-Ga$_2$O$_3$ films grown in long- and short-chamber MOCVD reactors before and after HT-XRD heating. (a,b) Surface topography of long- and short-chamber films measured before HT-XRD heating, respectively. (c,d) Corresponding surface morphologies after HT-XRD heating.

Figure 5 summarizes and benchmarks the crystalline quality and surface morphology of β-Ga$_2$O$_3$ films grown on various substrates by comparing FWHM and RMS roughness as functions of growth temperature and growth pressure. Figures 5(a) and 5(b) present the temperature dependence, while Figs. 5(c) and 5(d) show the corresponding pressure dependence.

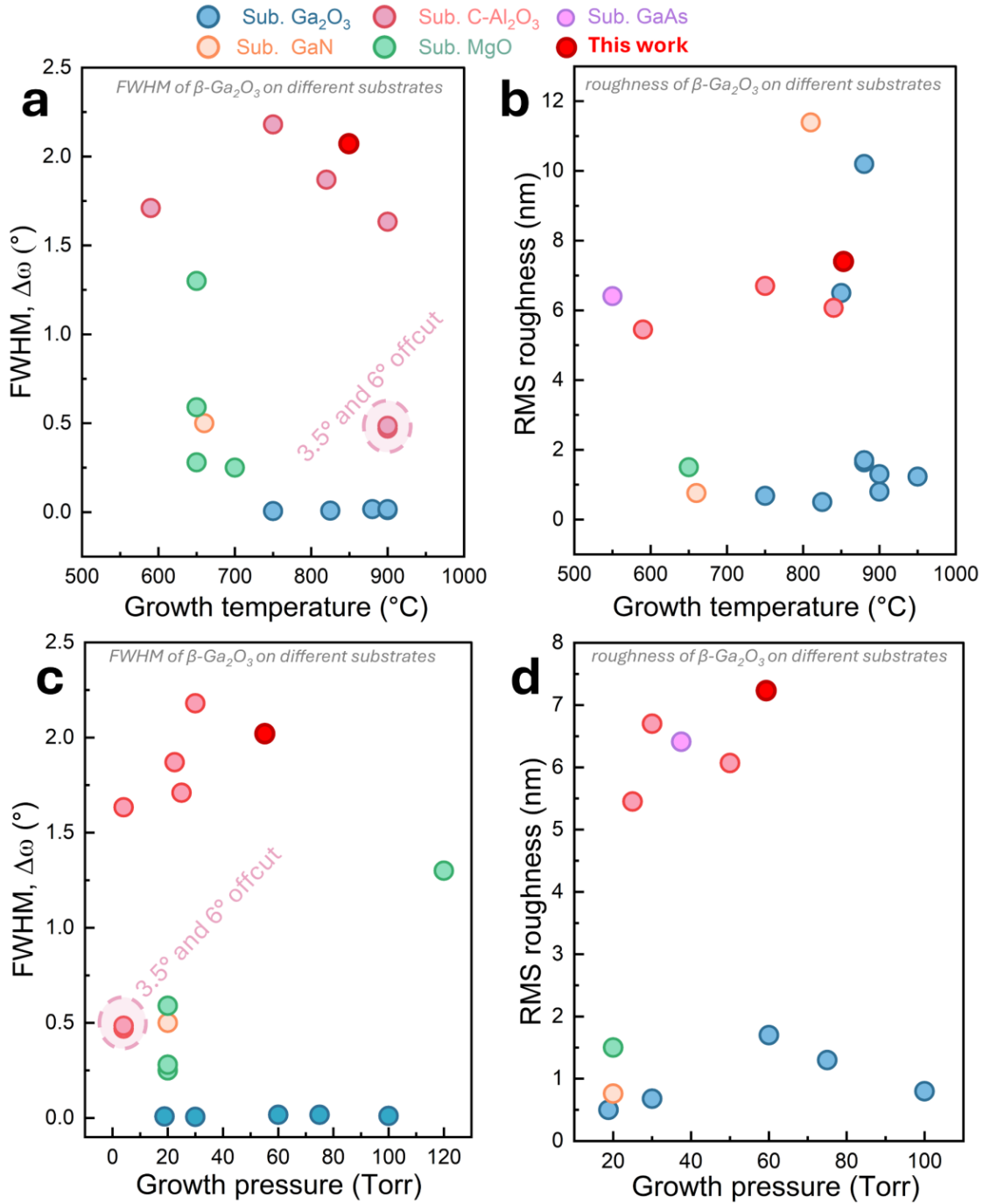

**FIG. 5.** Benchmarking of β-Ga$_2$O$_3$ film quality grown on different substrates.[16–20,22–25,35–44] (a,c) FWHM and (b,d) RMS roughness as functions of growth temperature and growth pressure for β-Ga$_2$O$_3$ films reported on β-Ga$_2$O$_3$, c-plane Al$_2$O$_3$, GaN, MgO, and GaAs substrates. The red star highlights the results achieved in this work.

As shown in Figs. 5(a) and 5(c), films grown on native β-Ga$_2$O$_3$ substrates consistently exhibit the lowest FWHM values across a wide range of growth conditions, reflecting the advantage of homoepitaxial growth. In contrast, heteroepitaxial growth on c-plane Al$_2$O$_3$, GaN, and MgO generally results in larger FWHM values and narrower growth windows, indicating increased mosaicity and strain associated with lattice and symmetry mismatch. Notably, the data point for this work (red star) falls among the lowest FWHM values reported on non-native substrates, demonstrating improved crystalline quality under optimized growth conditions. Figures 5(b) and 5(d) compare the trends in surface roughness. Films grown on β-Ga$_2$O$_3$ substrates maintain low RMS roughness over broad temperature and pressure ranges, while heteroepitaxial films exhibit higher roughness and stronger sensitivity to growth conditions. The RMS roughness achieved in this work is comparable to, or lower than, most reported values on c-plane Al$_2$O$_3$, highlighting the effectiveness of the growth strategy in simultaneously achieving low mosaic spread and smooth surface morphology. Overall, this benchmarking analysis places the present results within the context of existing reports and demonstrates that careful optimization of growth parameters enables β-Ga$_2$O$_3$ films with competitive crystalline and morphological quality on heteroepitaxial substrates.

In this work, we systematically examined the influence of reactor ceiling heights on the crystallinity evolution of β-Ga$_2$O$_3$ films grown by showerhead MOCVD by directly comparing long- and short-chamber configurations. Structural analyses reveal that differences in chamber height alter precursor transport and growth conditions, thereby establishing distinct as-grown structural and morphological states. During *in situ* high-temperature XRD heating from 25 °C to 1100 °C, both films exhibit a consistent redshift of the β-Ga$_2$O$_3$ (–201) diffraction peak, indicating a similar thermally driven strain response independent of reactor height. However, quantitative rocking-curve analysis shows that the short-chamber film achieves a smaller minimum FWHM of 2.03°, compared to 2.72° for the long-chamber film, reflecting differences in mosaic alignment established during growth. Phi-scan measurements confirm that these variations are not substrate-induced, isolating reactor heights as a contributing factor. Atomic force microscopy further shows that the short-chamber-grown films have smoother surfaces, with RMS roughness values of ~7.7 nm before and ~7.3 nm after HT-XRD heating, whereas the long-chamber-grown films exhibit higher roughness, decreasing from ~19.3 nm to ~12.3 nm after HT-XRD heating. Collectively, these results indicate that reactor height influences the initial crystalline and morphological template of $\beta$-$Ga_2O_3$ films and modulates their subsequent elevated-temperature structural response. These findings provide practical insight into how MOCVD reactor design can be leveraged to optimize $\beta$-$Ga_2O_3$ film quality for ultra-wide-bandgap device applications.

## AUTHOR DECLARATIONS

### Conflict of Interest

The authors have no conflicts to disclose.

### Author Contributions

**Imteaz Rahaman**: Data curation (equal); Formal analysis (lead); Investigation (lead); Methodology (equal), Writing-original draft (lead). **Botong Li**: Data curation (equal); Formal analysis (supporting); Investigation (supporting); Writing – review & editing (supporting). **Bobby Duersch**: Data curation (supporting); Formal analysis (supporting). **Hunter D. Ellis**: Writing – review & editing (supporting). **Kathy Anderson**: Methodology (equal), **Kai Fu**: Conceptualization (lead), Supervision (lead), Project administration (lead), Resources (equal), Writing – review & editing (lead).

**ACKNOWLEDGEMENT**

The authors acknowledge support from the University of Utah start-up fund. This work used the Nanofab EMSAL shared facilities of the Micron Technology Foundation Inc. Microscopy Suite, sponsored by the John and Marcia Price College of Engineering, Health Sciences Center, and the Office of the Vice President for Research. In addition, it utilized the University of Utah Nanofab shared facilities, which are supported in part by the MRSEC Program of the NSF under Award No. DMR-112125. Acquisition of the Bruker D8 Discover system was made possible by the Air Force Office of Scientific Research under project number FA9550-21-1-0293.

**DATA AVAILABILITY**

The data that supports the findings of this study are available from the corresponding authors upon reasonable request.